\documentclass[twocolumn,showpacs,preprintnumbers,amsmath,amssymb]{revtex4}
\pdfoutput=1
\usepackage[T1]{fontenc}
\usepackage{graphicx}
\usepackage{hyperref}
\begin{document}
\title{Probing viscoelastic properties of a thin polymer film sheared between a beads layer and quartz crystal resonator\\}
\author{J. L\'{e}opold\`{e}s}
\email{julien.leopoldes@univ-mlv.fr}
\author{X.P. Jia}
\email{jia@univ-mlv.fr}
 \affiliation{Laboratoire de Physique des Mat\'{e}riaux divis\'{e}s et Interfaces, UMR 8108 du CNRS, Universit\'{e} de 
Marne la Vall\'{e}e, 
Cit\'{e}e Descartes,  5 Bd Descartes, 77454 Marne la Vall\'{e}e cedex 2, France}
\affiliation{}%
\date{\today}

\begin{abstract}
We report measurements of viscoelastic properties of thin polymer films of 10-100 nm at the MHz range. These thin films are confined between a quartz crystal resonator and a millimetric bead layer, producing an increase of both resonance frequency and dissipation of the quartz resonator. The shear modulus and dynamic viscosity of thin films extracted from these measurements are consistent with the bulk values of the polymer. This modified quartz resonator provides an easily realizable and effective tool for probing the rheological properties of thin films at ambient environment.
\end{abstract}

\maketitle
From wet sand to the eye cornea, liquid systems confined into small volumes are ubiquitous in nature and are known to alter friction and adhesion at a solid-solid interface ~\cite{isra1,robbins1,bocquet1,brunet1}. Moreover the mechanical properties and stability of thin films is of paramount importance for a number of applications requiring specific nanometric coatings such as optical reflectors or dielectric stacks. It is then natural to ask weather the properties of confined liquids are similar to their bulk counterpart. 
Conventional mechanical testing is not adapted for thin films investigation and specific metrology is needed. Naturally occurring instabilities such as wrinkling or dewetting provides valuable information about rheological and mechanical properties of thin polymer films ~\cite{harrisson1,fret1}. Oscillatory interfacial rheology conducted with a tribometer ~\cite{bureau1} and the Surface Force Apparatus (SFA) ~\cite{granick1,isra2} offers a complementary description of thin films over several decades of frequency. Local probes analysis performed with Atomic Force Microscopy confirms the result obtained with the SFA ~\cite{kappl1}. Note that the rheological behaviour at high frequency can not be investigated with these techniques. However it is important for applications such as hard disk drives.

Acoustic wave devices have proven to be suitable for the high frequency investigation ~\cite{frye1,royer1}. Among others, quartz crystal (AT-cut) resonators operating in shear mode at ultrasonic frequency from $1-100$ MHz have been the most widely used to monitor the viscoelastic properties of thick films ($h > 1 \mu$m) adsorbed on their surfaces ~\cite{frye1}. However for thin films, i.e. $h \leq 100$ nm, no significant shear strain is produced inside it because the film is located at an antinode of the standing wave. In such a case, the shift of resonance frequency is only a function of the mass alone and the acoustic properties of the film may be ignored: the quartz crystal behaves simply as microbalance (QCM). Recently, several configurations have been proposed to determine the shear modulus of thin films from the frequency shifts of quartz resonators, including the build-up of composite resonators. For example, coating the film of interest with second overlayer (sandwich configuration) allows enhancing the shear stress and characterizing its rheological properties down to nanometric thicknesses ~\cite{johann2}.
In this Letter, we describe a new approach to probe the viscoelastic properties of thin films by using a conventional QCM. To this end, we deposit gently a layer of spherical bead (glass) on the top of a film of thickness $h \sim 10 - 100$ nm coated on the surface of a quartz crystal. The resulting shifts of resonance frequency and inverse of quality factor can be readily related directly to the elastic modulus and viscous dissipation of the film. 
\begin{figure}[htbp]
	\centering
		\includegraphics[width=7cm]{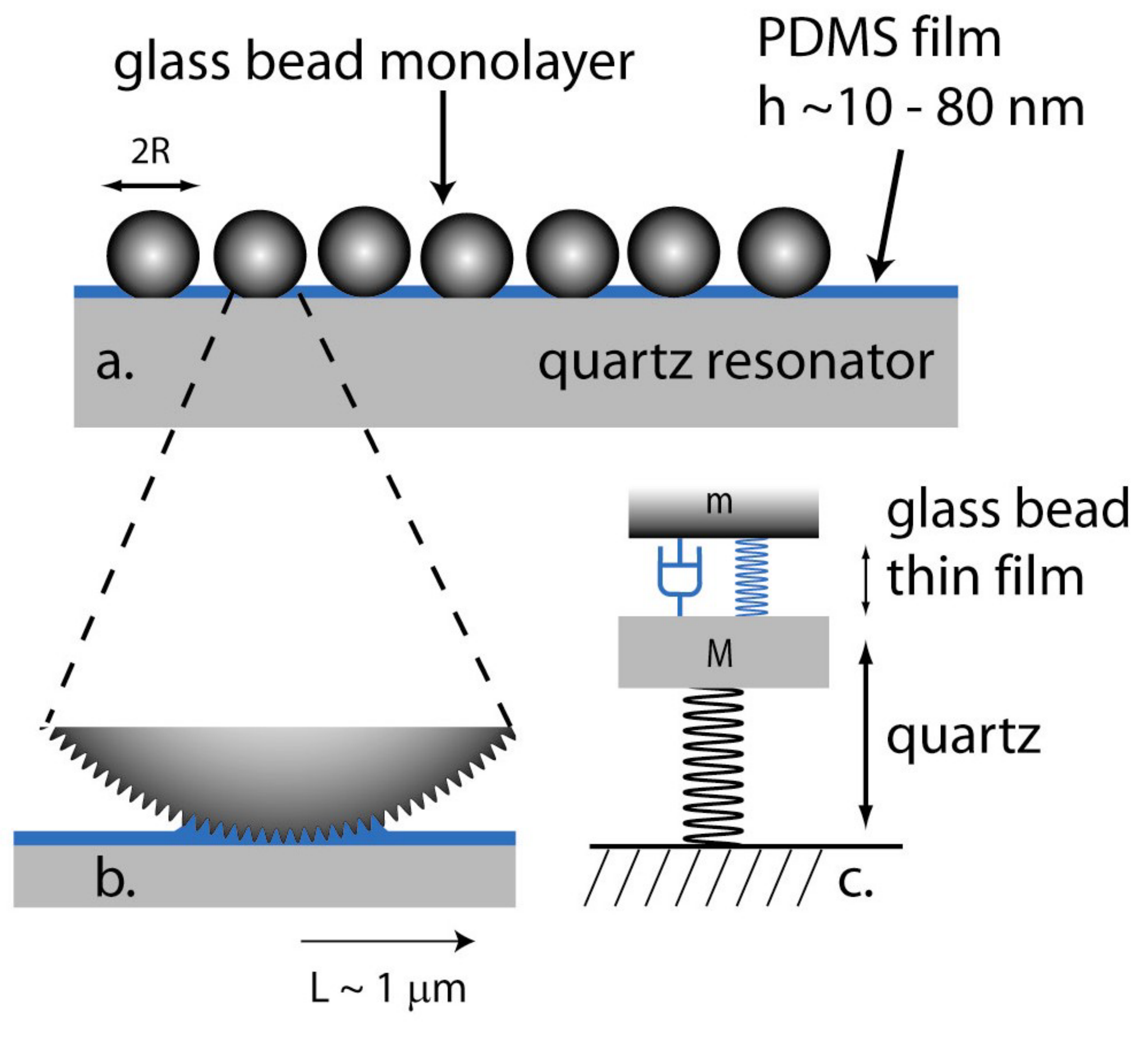}
	\caption{Sketch of the experiment and modeling of the system}
	\label{fig:schema1}
\end{figure}
An entangled polymeric thin film (polydimethylsiloxane, $Mw \sim 90000$) is deposited on a quartz resonator of $f_{0}=\omega_{0}/2\pi\sim 5$MHz (Maktex) (Figure \ref{fig:schema1}). The surface of the crystal is polished and gold-coated with a roughness of about $2$ nm. This quartz is first cleaned by snow jet, and then by oxygen plasma during 10 minutes. The film under study is deposited by spin coating directly onto the quartz from a heptane solution. The film-coated quartz is placed in a home-made cell and allowed to stabilise at $40^{\circ}$C during 1 hour.
We use an impedance analyser (Solartron 1260A) to measure the admittance spectrum of the quartz resonator at different stages of the sample preparation. As shown in Figure \ref{fig:figure2} (inset), the resonance frequency $f_{0}$ is determined by curve fitting with a precision of  $\Delta f_{min} = \pm 1 $Hz and the quality factor is obtained from $Q_{0}=f_{0}/\Delta f$ ($\sim 10^{5}$) where  $\Delta f$ is the width of resonance peak. The deposition of the film lowers the resonance frequency of the crystal; the measurement of such a frequency shift allows determining the thickness of the film ~\cite{frye1}. Slight fluctuations of the resonance peak amplitude are observed when mounting the quartz into the measurement cell from one experimental run to another, but the downward shifts of resonance frequency are very reproducible. 
In order to study the viscoelastic properties of a deposited thin film beyond the QCM application, we cover the film-quartz system with a monolayer of glass beads of diameter $2R \sim ˜ 400 \mu$m (Figure \ref{fig:schema1}a). These beads for abrasive use (from Centraver) are plasma cleaned and have a surface roughness of about 100 nm. Covering the film with glass beads results in a totally different behaviour than usual mass loading (Figure \ref{fig:figure2}a). We observe an increase of the resonance frequency  $\Delta f^{+}$ after the deposition of beads, and a decrease of resonance amplitude together with peak broadening. These results clearly indicate elastic enhancement and dissipation increase of the quartz resonator. Moreover both frequency shift  $\Delta f^{+}$ and energy dissipation  $\Delta Q^{-1}$ evolve with time in a similar way roughly according to a power law. This suggests a linear relationship between  $\Delta Q^{-1}$ and $\Delta f^{+}/f_{0}$ as plotted in Figure \ref{fig:Graph5} for various thicknesses.
\begin{figure}
	\centering
		\includegraphics[width=7cm]{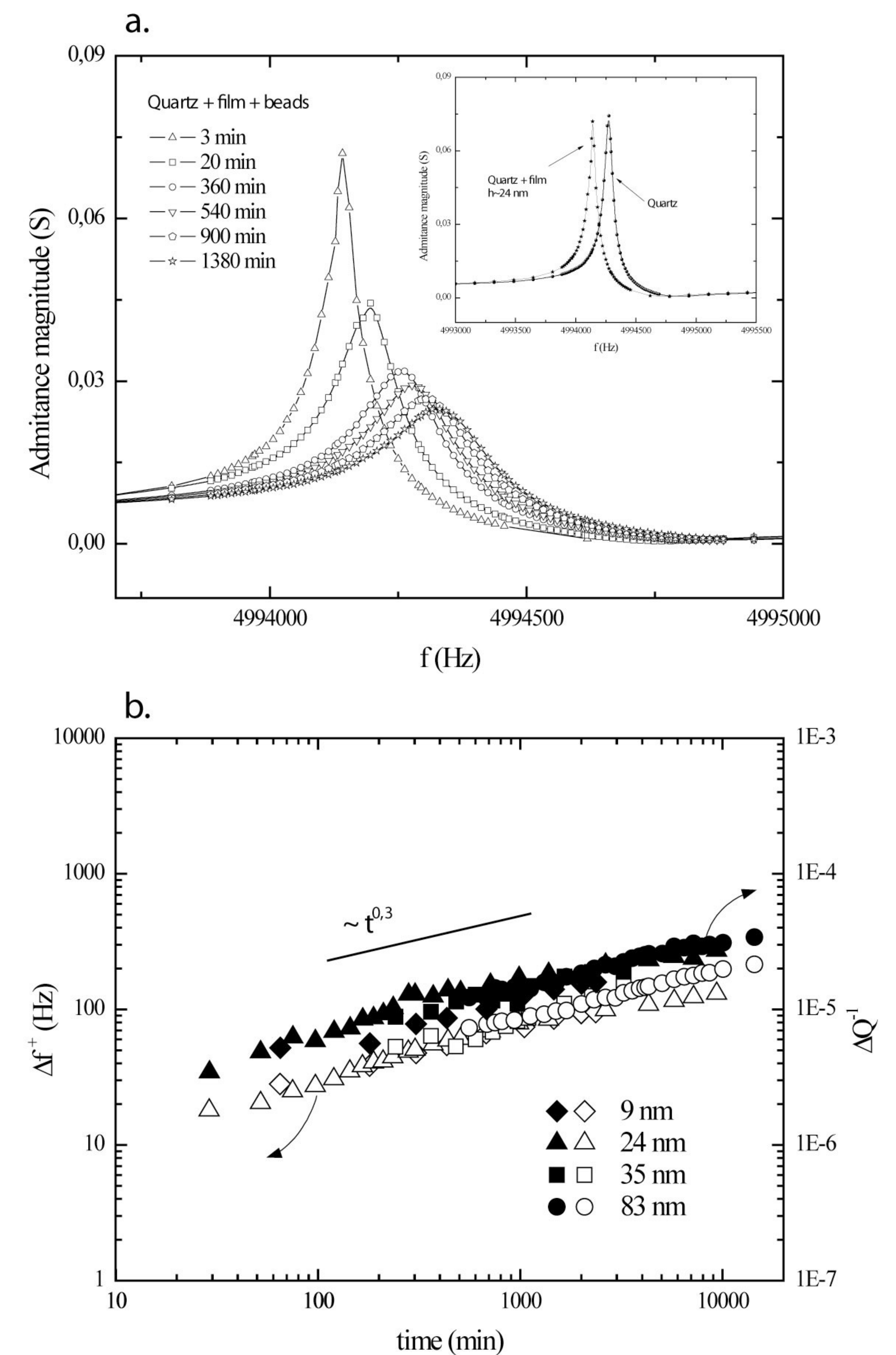}
	\caption{a. Resonance peak at various stages of the experiment. b. Ageing of resonance frequency}
	\label{fig:figure2}
\end{figure}
Analysis and modelling: To gain physical insight into our resonance measurements, we model the quartz/film/beads assembly as a pair of coupled oscillators (Figure \ref{fig:schema1} b,c). As in a previous work ~\cite{dybwad1}, the quartz resonator can be viewed as an effective masse $M$ attached to a spring of shear stiffness $K$, determined from the fundamental resonance frequency $K = M\omega_{0}^2$. The glass bead of mass $m$ is attached to the resonator $M$ via the adsorbed film. This film is modelled by a Kelvin-Voigt element with spring of shear stiffness $k$ and damping constant $G"$ (see discussions below). No-slip boundary conditions are assumed here between both the film-bead and quartz-film interfaces, which are ensured by the surface roughness of beads and the strong adsorption of polymer films considered here.
Two eigenmodes exist for such coupled oscillators system with a natural frequency either superior ($\omega^{+}$) or inferior ($\omega^{-}$) to the quartz frequency $\omega_{0}$. The  $\omega^{-}$ - mode corresponds to an \textit{in-phase} motion between $m$ and $M$ giving access to the mass deposition by beads (not discussed here), while the  $\omega^{+}$ - mode corresponds to an out-of-phase motion. The $\omega^{+}$ -mode is detected by the quartz crystal (Figure \ref{fig:figure2}a) and to first-order approximation in $k$, the associated frequency shift $\Delta \omega^{+}$ ($= \omega^{+}-\omega_{0}$) induced by the layer of beads reads:
\begin{equation}
\Delta \omega^{+} \approx N_{b}k/2\sqrt{MK}
\label{eq1}
\end{equation}
Here $N_{b} \sim  300$ is the number of beads effectively covering the quartz electrode; the oscillation of all beads is assumed to be identical. The determination of $\omega^{+}$ enables us to characterize the elastic enhancement (stiffening) and properties of the adsorbed or bonded films. The minimum shear stiffness $k_{min}$ that can be measured with this experiment is determined by $\Delta f^{+}_{min}$ leading to $k_{min} \sim 100 $N/m for $K = 3 .10^{10}$ N/m and $M = 3 .10^{-5}$ kg. 
Moreover, the relative motion between the beads and the quartz resonator, inherent to the    mode, enhances strongly the shear strain in the adsorbed film and induces an interfacial dissipation which shall be detectable with the quartz resonator. It has been shown previously that the interfacial dissipation between two dry rough solid surfaces is governed by the interplay of a frictional loss and an interfacial viscoelastic one~\cite{brunet1, baumb1}. However in the presence of wetting films as the case here, the viscous loss appears predominant over the other contributions to the energy dissipation~\cite{brunet1, crassous1}. To characterize such a loss, we calculate the dissipated energy per cycle of oscillation by shearing a thin film between a sphere and a flat surface (Figure \ref{fig:schema1}b), $\Delta W_{film}\approx 2\pi(\omega^{+}\eta)(AL^{2}/2h)U^{2}$. Here $\eta$   is the film viscosity, U ($\sim 1$nm) is the vibration amplitude of the quartz and $A = ( 16/5 lg(2R/h)$ is a geometrical constant. In terms of the inverse of quality factor  $\Delta Q^{-1} = (2\pi)^{-1}(\Delta W_{film}/W_{q})$ where $W_{q} = (1/2) KU^{2}$ is the stored energy in the quartz, the additional dissipated energy is written 
as ( $\omega^{+}\approx\omega_{0}$),
\begin{equation}
\Delta Q^{-1}=AN_{b}G"L^{2}/ 2hK
\label{eq2}
\end{equation}
where $G" = \omega_{0}\eta$ and $L$ is the effective radius of the contact. Measurements of such $\Delta Q^{-1}$ ($\sim 10^{-5}$) allow us to determine the dynamic viscosity of a thin polymer film down to a thickness of 10 nm (Figure \ref{fig:figure2}b). 

We now focus on the possible mechanisms responsible for the elastic stiffness k observed in our experiments. As shown in Figure \ref{fig:figure2}b, the frequency shift for every measurement is about $\Delta f^{+}\sim 100$ Hz, which corresponds to stiffness of the order of $k \sim 10^{3}$ N/m for one bead contact. Dybwad ~\cite{dybwad1} previously reported such a stiffness magnitude between Au spheres on Au coated quartz resonator, originating from Van der Waals bonds. This cohesion is not observed in our resonance measurements when the glass beads are deposited on the bare quartz crystal. This may be related to the roughness of the beads, which can reduce significantly the contact stiffness of Hertz-Mindlin between glass spheres and the quartz resonator, from $k_{HM} = 3.10^{4}$ N/m ($>> k_{min}$) on smooth interface to $k_{MCI} = 1$ N/m on rough (multi-contact) interface ~\cite{baumb1}. This latter value is too low to be detected with our present apparatus indeed.
\begin{figure}
	\centering
		\includegraphics[width=7cm]{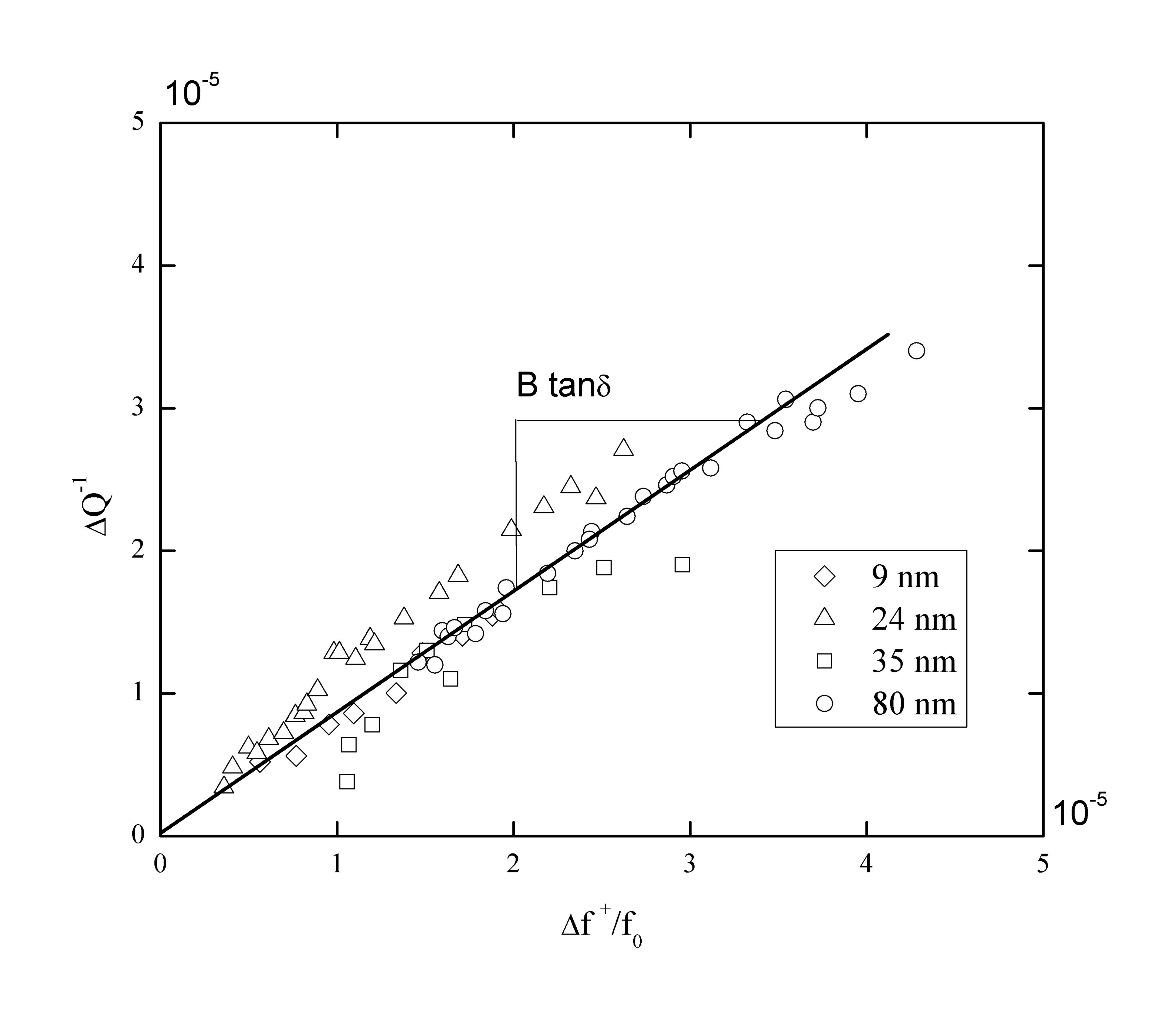}
	\caption{Elasticity vs dissipation}
	\label{fig:Graph5}
\end{figure}
In the presence of a wetting film, the capillary force $F_{c}$ increases the normal loading on the glass bead (Figure \ref{fig:figure2}b) and could stiffen the above contact stiffness ~\cite{johann1}. An estimation with the liquid-air surface tension $\gamma$ yields a force strength on rough surface ~\cite{hasley1}, $F_{c} \sim 10^{-7} $N, which is less than the weight of bead $F_{b} \sim 10^{-6} $N; this implies that no capillary effect is expected on the contact stiffness in our work. However, the wetting film may introduce the elastic stiffening via another mechanism, related to the stiffness of larger capillary bridges $k_{c}$. With a meniscus formed between a smooth sphere and a flat surface, a stiffness $k_{c} \sim 2\pi L^{2}\gamma /h^{2}$ of the order of $10^{3}$N/m is expected ~\cite{crassous1}. This mechanism predicts a dependence of $k_{c}$ on thickness that is not evidenced experimentally (Figure \ref{fig:figure2}b). This is possibly due to the effective coupling and complex wetting of mechanisms of rough surfaces, somehow characterized by $N_{b}$ in eqs (1) and (2) fluctuating from one experiment to another. Representing the data such as  $\Delta Q^{-1}$ versus  $\Delta f^{+}/f_{0}$ shall allow overcoming this caveat (Figure \ref{fig:Graph5}). However, the resulting thickness dependence $\Delta Q^{-1} \sim 1/h (\Delta f^{+}/f_{0})$ is not detected either.

We propose here an interfacial mechanism based on the elastic behaviour of the adsorbed films. At the MHz range, polydimethylsiloxane is in the Rouse regime as shown by the characteristic frequencies of a Kuhn monomer $\tau_{\kappa}^{-1}=k_{B}T/\zeta b^{2}\sim10^{8}$ Hz and of an entanglement strand $\tau_{e}^{-1}=\tau_{\kappa}^{-1} N_{e}^{-2}\sim10^{5}$ Hz. Here, $\zeta \sim 10^{-11}$ kg/s~\cite{ferry} is the friction coefficient of a monomer,  $k_{B}$ is the Boltzmann constant, $T$ the temperature and $b$ the lenght of a Kuhn monomer). At such high-frequency range the polymeric layer would provide a stiffness $k_{e}\sim \pi L^{2}G'/h$. Comparison between our experiment and this model leads to $G' \sim 5 $MPa and $G" \sim 0.1 $MPa (eq. 2), which agree well with those expected for bulk polydimethylsiloxane ~\cite{rahalker}. This picture also provides a simple relationship between the polymer loss angle $tan \delta =G"/G'$ and the plot of  $\Delta Q^{-1}$ vs $\Delta f^{+}/f_{0}$. Indeed, combining eqs 1 and 2 yields
\begin{equation}
\Delta Q^{-1} \approx B tan\delta \Delta f^{+}/f_{0}
\label{eq3}
\end{equation}
with $B \sim 60$, which is consistent with the scaling behaviour observed experimentally in Figure \ref{fig:Graph5} and indeed independent of film thickness h. We thus conclude that the viscoelasticity of the polymer film appears dominant and responsible for the elastic ($\Delta f^{+}$) and dissipative ($\Delta Q^{-1}$) responses of our quartz resonator. 

Our experiment shows no dramatic change of the viscoelastic response of thin films. As mentioned earlier, polymer films may have unusual properties when confined into narrow gaps. For example for film thicknesses $h \sim 10 R_{g}$ ($R_{g}$ is the radius of giration) the low frequency dynamic moduli of PDMS show a non monotonic dependence on film thickness and the terminal zone shifts progressively to lower frequencies ~\cite{isra3}. No rheological data is however available for the high frequency region. In this work the thickness of the films ranges from $1-10 R_{g}$ ($R_{g}\sim 7$ nm), but no unusual behaviour is detected. This could be expected given the very local probe provided by the present method. At such high frequency, only Rouse modes are probed (size < nm) and no confinement effect is expected until the thickness of the film reaches the characteristic size of those modes. The generality of our observation has to be confirmed by additional measurements down to smaller thicknesses.

We now turn to the ageing phenomenon observed both with the elastic stiffening ($\Delta f^{+}$) and the energy dissipation ($\Delta Q^{-1}$) (Figure \ref{fig:figure2}b). In a previous work ~\cite{bocquet1}, Bocquet et al. described a thermally activated formation of liquid bridges between rough glass beads, which is responsible for the logarithmic ageing of capillary cohesion in a granular medium exposed to water vapour. In our experiments, both $\Delta f^{+}$ and $\Delta Q^{-1}$ are seen to increase with time following a power law $\sim t^{0.3}$. This could be related to a progressive wetting of the glass bead by the polydimethylsiloxane thin film. Indeed, we observe by optical microscopy the increase of contact radius L with time, revealing an evolution roughly as $L ~ t^{0.16}$ (not shown here). This result can be understood by the following scaling argument. During a wetting process, some polymer liquids must drain from the thin film to the free surface of the beads. If $L$ is the lateral extent of the contact between the sphere and the quartz, a Poiseuille flow yields $\frac{dL}{dt}=\frac{h^{2}\Delta P}{\eta L}$ where $\Delta P$ is the Laplace pressure. This leads to $L\sim(\gamma h^{4}L^{2}t/\eta)^{1/7}$. Assuming a constant average thickness of the film, this prediction agrees reasonably well with the observed time evolution of $\Delta f^{+}$ and $\Delta Q^{-1}$. Note that the surface roughness of the beads is not taken into account in this analysis. This problem is beyond the scope of this work and will be treated in the future. 

In summary, we have developed a new ultrasonic method for measuring high-frequency shear modulus and dissipation of thin films down to 10 nm in thickness. Our results indicate that the viscoelastic properties of such a polymer film are not quantitatively different from those of the bulk. We believe that by using beads of different surface properties and controlling ambient conditions, this ultrasonic method provides a promising tool for exploiting the confinement effects of nanometric films, the interfacial dynamics and the wetting phenomena at various boundaries~\cite{robbins1,krim1}.
 
\acknowledgments{We wish to thank J. Laurent for his assistance to impedance measurements and the design of the cell, D. Hautemayou and H. Sizun for the cell realization, and Y. Leprince for useful discussions.}
 
\bibliographystyle{unsrt}

\end{document}